\documentclass[aip,reprint]{revtex4-1}
\usepackage{graphicx,subfig}  
\usepackage{dcolumn}   
\usepackage{bm}        
\usepackage{amssymb}   

\begin{document}
\title{Anomalous magneto-structural behavior of MnBi explained: \\ a path towards an improved permanent magnet}
\author{N.A. Zarkevich} \email{zarkev@ameslab.gov}
\author{L.-L. Wang}
\affiliation{The Ames Laboratory, U.S. Department of Energy, Ames, Iowa 50011 USA}
\author{D.D. Johnson}
\affiliation{The Ames Laboratory, U.S. Department of Energy, Ames, Iowa 50011 USA}
\affiliation{Materials Science and Engineering, Iowa State University, Ames, Iowa 50011 USA}

\date{\today}

\begin{abstract}
Low-temperature MnBi (hexagonal NiAs phase) exhibits anomalies in the lattice constants ($a$,\,$c$) and bulk elastic modulus (B) below 100 K, spin reorientation and magnetic susceptibility maximum near $90\,$K, and, importantly for high-temperature magnetic applications, an increasing coercivity (unique to MnBi) above $180\,$K. 
We calculate the total energy and magneto-anisotropy energy (MAE) versus ($a$,\,$c$) using DFT+U methods. 
We reproduce and explain all the above anomalies.  
We predict that coercivity and MAE increase due to increasing $a$, suggesting means to improve MnBi permanent magnets.
\end{abstract}


\maketitle

{\par}
MnBi in its low-temperature phase (LTP) has one of the most extraordinary magnetic properties among ferromagnetic materials. \cite{Heusler1904,Arrivaut1905,JAP23p1207y1952,ProcPhysSocB67p290y1954,Shchukarev1961,JPhysSocJapan16p2187y1961,ActaChemScand21p1543y1967,JETPLetters30p333y1979,JAC317p297y2001,APL79p1846y2001,JPhysCM14p6509y2002,JAP91p7866y2002,JJapanInstMetals73p40y2009,SciTechnolAdvMater9p024204y2008,APL99p082505y2011,JAC509pL78y2011,JMMM324p1887y2012,JMMM115p66y1992,PSSA34p553y1976,PSSA30p251y1975,JAP43p2358y1974,JAP39p5471y1968} 
Uniquely, its coercivity increases with temperature (T), and its value is larger than that of Nd$_2$Fe$_{14}$B above $423\,$K, making it potentially an excellent permanent magnet for higher-temperature applications. 
MnBi does not contain critical rare-earth elements and, thus, it has a potential for technological impact. 
If magnetic anisotropy energy (MAE) is better controlled and tuned, use of MnBi magnets could be broadened.
Below we provide theoretical explanation for the long-standing experimental puzzles in the measured coercivity, spin orientation, lattice constants, and bulk modulus of MnBi. We also suggest a means to further increase the MAE.

\begin{figure}[b]
\includegraphics[width=82mm]{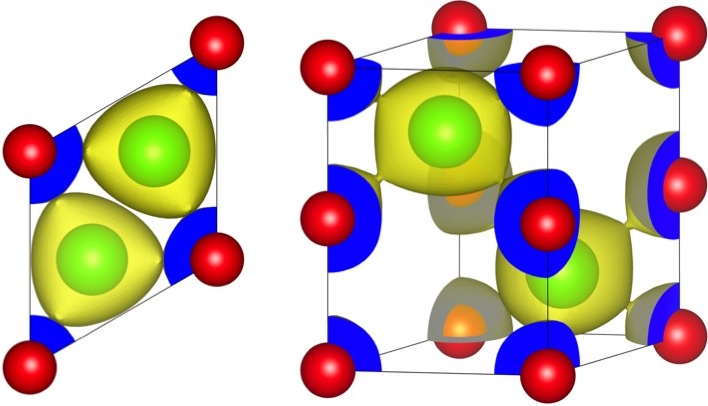}
\caption{\label{fig1} MnBi hexagonal structure ($hP4$, P63/mmc, No.194), with $0.0323\, e/\mbox{\AA}^3$ charge density isosurfaces. (0001) projection (left), and primitive unit cell (right) with two Mn (red) and two Bi (green) atoms.} 
\end{figure}

{\par }
Despite its simple NiAs hexagonal structure (Fig.~\ref{fig1}), stable below $628\,$K, \cite{Shchukarev1961,JJapanInstMetals73p40y2009} MnBi exhibits several puzzling and unexplained behaviors versus T. \cite{JPhysCM14p6509y2002,JAP91p7866y2002,JJapanInstMetals73p40y2009,SciTechnolAdvMater9p024204y2008,APL99p082505y2011}
First, the lattice constant $a$ exhibits minimal thermal expansion below 70 K and then expands rapidly during the spin reorientation, while $c$ shows a chaotic zigzag behavior below $150\,$K. \cite{JAC317p297y2001,SciTechnolAdvMater9p024204y2008,APL99p082505y2011} 
Second, there is a measured  kink in the bulk modulus (B) near 39 GPa at $100\,$K. \cite{JETPLetters30p333y1979} 
Third, a spin reorientation is observed at $T_{SR} \approx 90\,$K, \cite{JETPLetters30p333y1979,JPhysCM14p6509y2002,JAP91p7866y2002} 
when the magnetization M(T) easy axis changes from in-plane to $c$-axis above $T_{SR}$. 
Next, coercivity is near zero at T$<$180$\,$K, and increases with T above 180$\,$K. 
Finally, above $628\,$K MnBi transforms to a high-T oP10 phase (stable between $613\,$K and $719\,$K) with M=0. \cite{ActaChemScand21p1543y1967}

\begin{figure}[t]
\includegraphics[width=80mm]{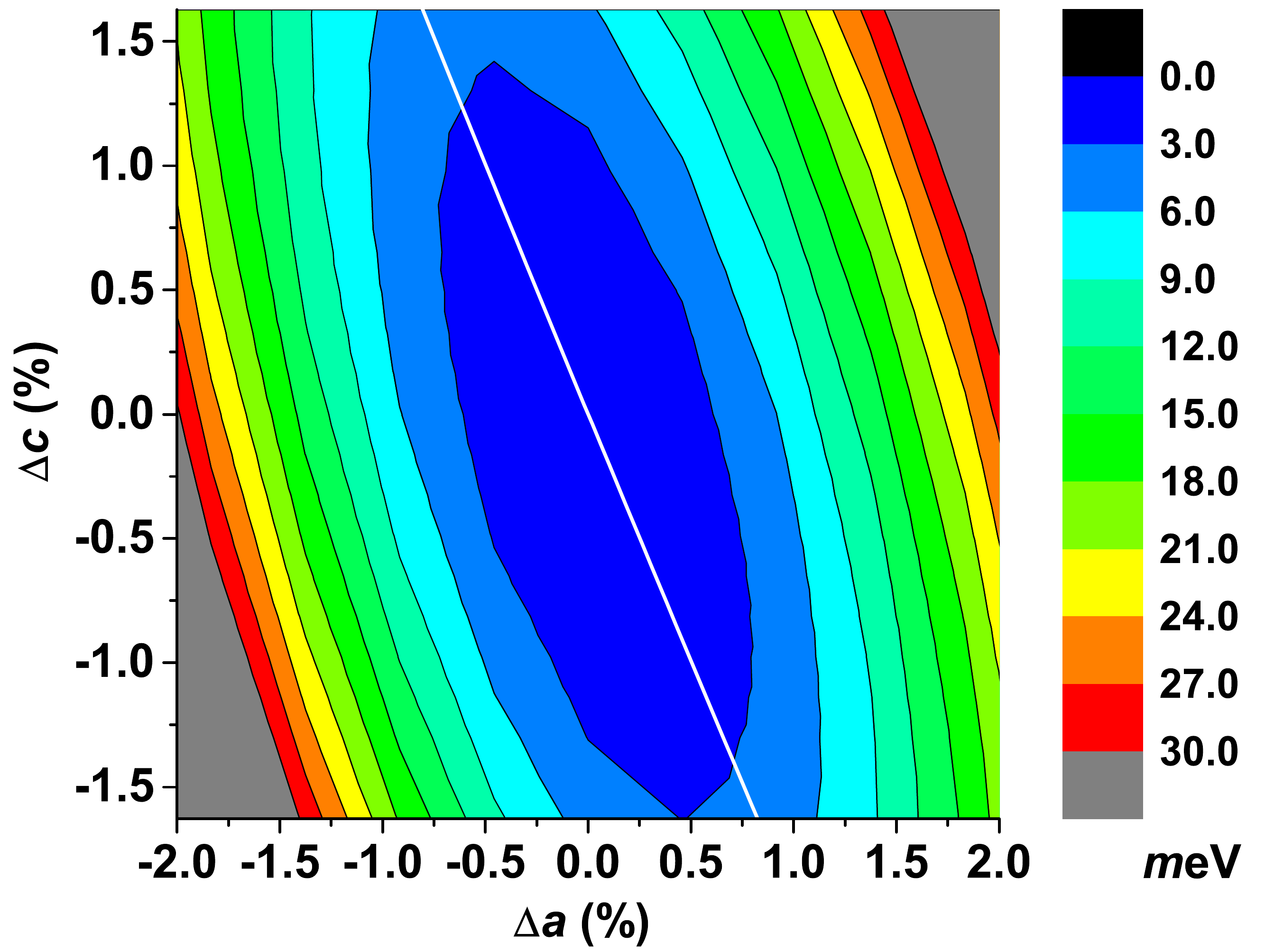}  \\
\caption{\label{figEnergy} Change in the total energy E at 0~K vs. ($\Delta a,\Delta c$), with $3\,me$V/cell (or $34.8\,$K) between contours.  Constant volume $(ca^2 \sqrt{3}/2)$ is the line through (0,0).   }
\end{figure}

\begin{figure}[t]
\includegraphics[width=80mm]{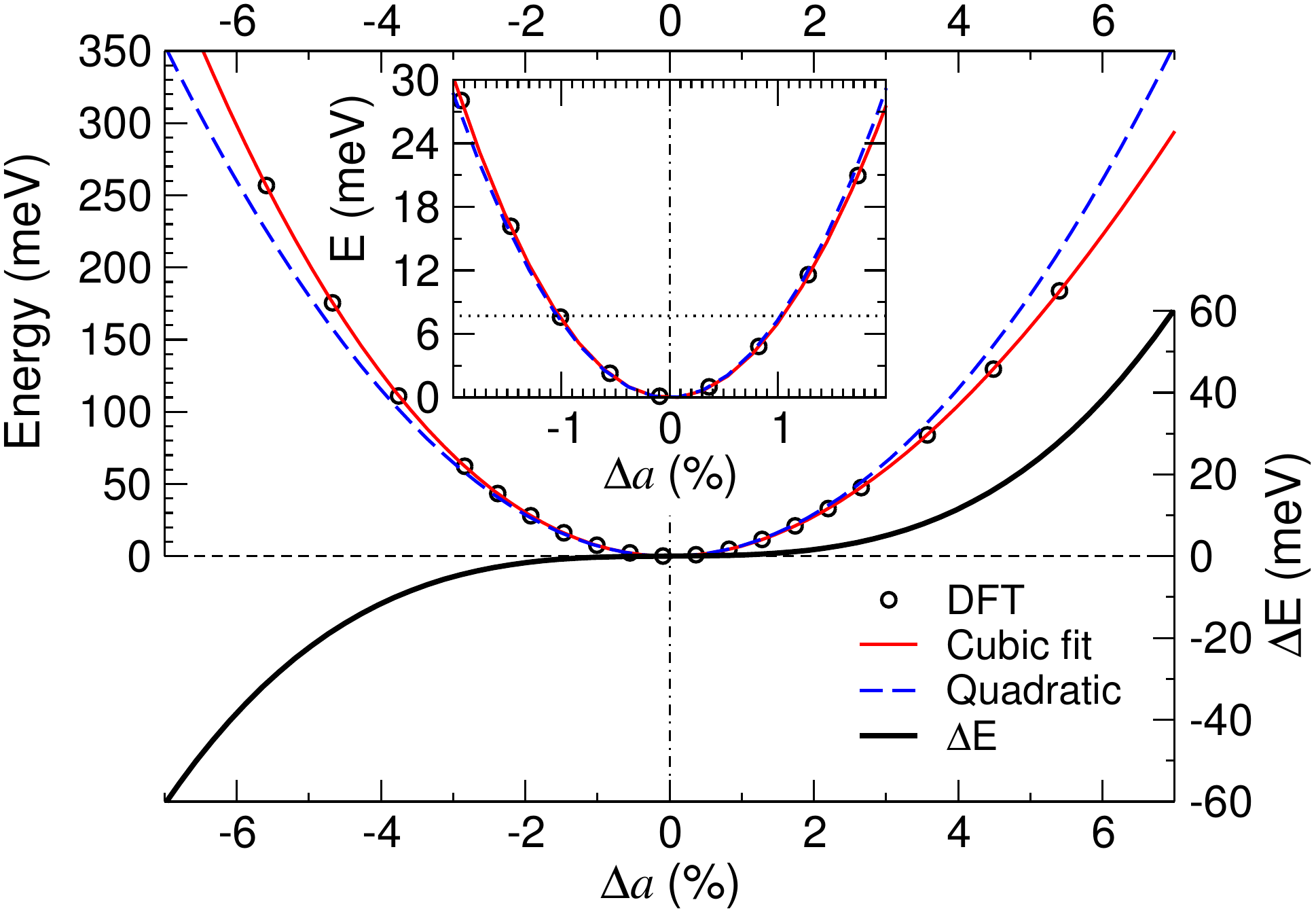}
\caption{\label{figEa} E$(\Delta a,\Delta c =0)$. GGA+U results (circles) with cubic (line) and quadratic (dashed) fits, and their difference $\Delta $E (black line, right scale).  
$\hbar \omega = 7.7\,m$eV ($90\,$K) is the horizontal dotted line in the inset.  }
\end{figure}

\begin{figure}[t]
\includegraphics[width=80mm]{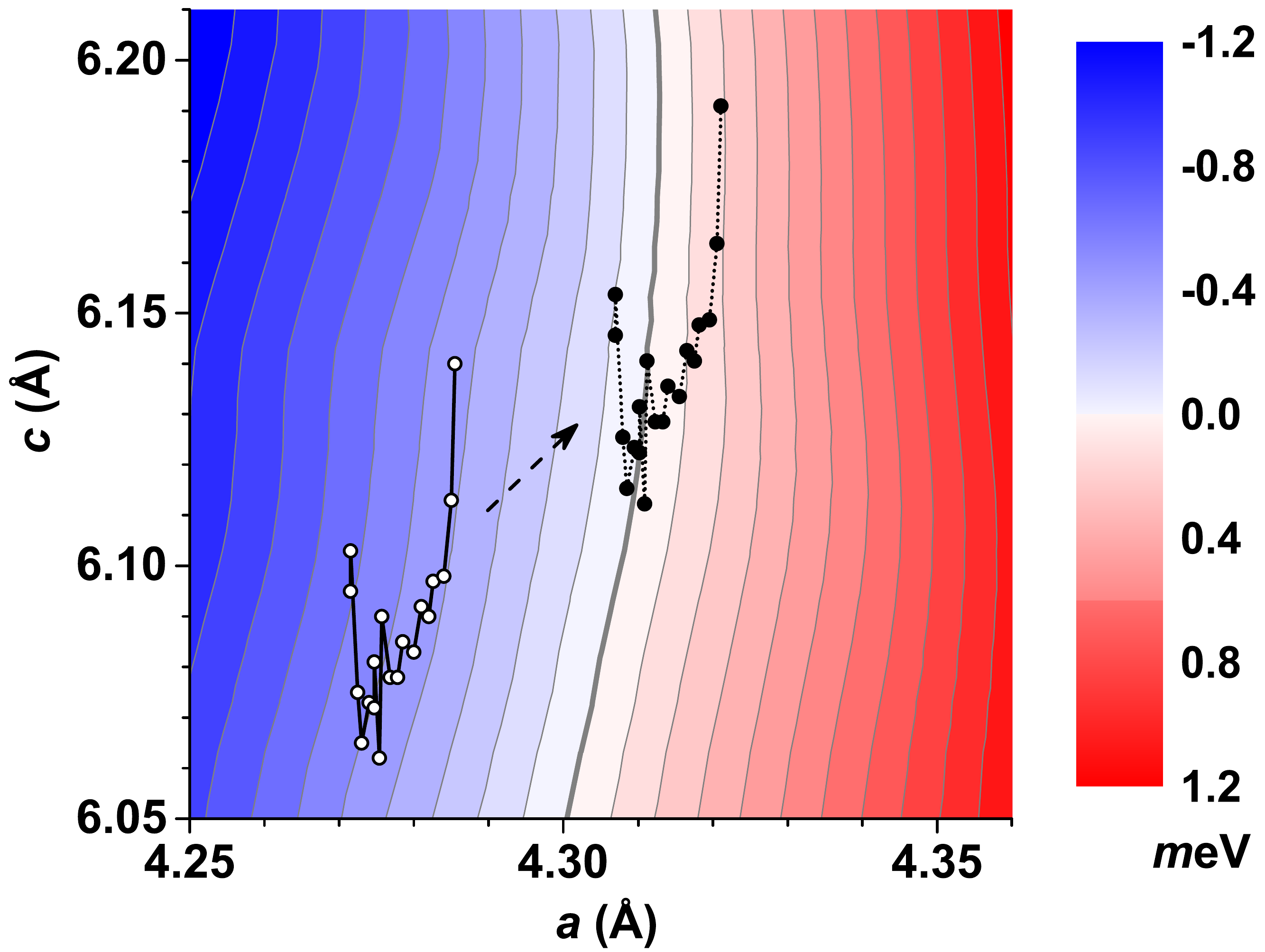}
\caption{\label{figMAE} MAE vs. ($a,\,c$) with $0.1~me$V/cell steps in contours from zero (grey line). Assessed data \cite{SciTechnolAdvMater9p024204y2008} (circles) is shifted by 0.83\% (filled circles), see text. }
\end{figure}

{\par}
We explain all these observations by examining  dependence of the calculated total energy (E) and MAE on the lattice geometry (Figs.~\ref{figEnergy},~\ref{figEa} and \ref{figMAE}). 
The total energy is anisotropic versus ($a,c$), like a ``flat-bottom canoe,'' and its asymmetry causes abnormal thermal expansion.
Due to the nature of the potential energy surface, the second derivative of the total energy with respect to volume 
is not monotonic, producing a kink in B$=V d^{2}E/dV^{2}$ near 39 GPa, whose origin can be traced to features in electronic density of states (DOS). 
Spin reorientation arises from a change of sign in MAE, which depends on increasing $a$, This suggests simple means to control MAE: by thermal expansion (observed), or by strain or alloying, e.g., coherent interfacing or doping. 
While temperature and strain affect mostly ($a,c$), doping can induce competing effects on MAE, some of which can be beneficial. Preliminary results suggest that doping with selected metals (Ni, Rh, Pd, Ir) increases MAE and coercivity and stabilize the spin orientation along $c$ at all temperatures.

{\par} {\em Computational method:}
We use a DFT+U method implemented in the Vienna {\it ab initio} simulation package (VASP). \cite{VASP1,VASP3}
We use $16 \times 16 \times 10$ Monkhorst-Pack $k$-point grid with the $\Gamma$-point, a $337.4$ eV plane-wave energy cutoff and $500.7~e$V augmentation charge cutoff, for both energy and magnetic anisotropy energy. 
A modified Broyden's method \cite{PRB38p12807y1988} is used for electronic self-consistency.
Bulk moduli are found from dependence of the total energy E$(a,c)$ on volume $V=ca^2 \sqrt{3}/2$. 
MAE is the energy difference with moments along $\langle1\bar{2}10\rangle$ and then $\langle0001\rangle$, i.e., E$[1\bar{2}10]-$E$[0001]$.
Generally, the MAE can be the order of $\mu e$V to $me$V; in MnBi for changes in $a$, pertinent to thermal lattice expansion effects, changes are order of $me$V.

{\par} 
We improve description of the electronic structure (and, hence, magnetization and lattice parameters) by combining the spin-polarized, generalized gradient approximation (GGA) \cite{PW91} with the rotationally invariant DFT+U formalism.\cite{LDAUTYPE2} 
GGA includes local value and gradient of the electron density $n=n_{\uparrow}+n_{\downarrow}$
and spin density $n_{\sigma}$ ($\sigma$=$\uparrow,\, \downarrow$) in the exchange-correlation functional 
$ E_{xc}^{GGA} [n_{\uparrow}, \nabla n_{\uparrow}, n_{\downarrow}, \nabla n_{\downarrow}] $.\cite{textbookGGA}
DFT+U corrects the total energy\cite{LDAUTYPE2} for presence of localized states, i.e.,
$E_{DFT+U} = E_{DFT} + \frac{1}{2} (\mbox{U}-\mbox{J}) \sum_{\sigma} (n_{m,\sigma} - n_{m,\sigma}^2)$,
where $n_{m,\sigma}$ is the occupation number of state $m$ ($m=2$ for $d$-orbital on Mn).
See textbook \cite{textbookLDAU} for more details.
After testing, we set (U$-$J)=2\,$e$V for correlated Mn $d$-electrons to better reproduce the measured ($a,c$) and M (Table~1). 
Note, a single U$-$J parameter cannot be adjusted to reproduce both ($a,c$) and M perfectly. 
At $0\,$K, we find $a$ ($4.363${\AA}) and $c$ ($6.123${\AA}) in good agreement with those measured \cite{JPhysCM14p6509y2002} at $10\,$K (Table~1), with an overestimate by 1.86\% and 0.21\%, respectively. 
The calculated M(0) is $3.96\, \mu_B$/MnBi (with site-projected moments of 4.231 and $-0.273\, \mu_B$ on Mn and Bi, respectively); it agrees with the extrapolated to $0\,$K values of 4.0 \cite{PhysRev104p607y1956}  and $3.95\, {\mu}_B$; \cite{PhysRev99p446y1955} or the measured values of $3.84 \pm 0.03\, \mu_B$ at $4.2\,$K; \cite{AIP18p1222y1974}   $4.18\, {\mu}_B$ at $10\,$K, or $3.60\, {\mu}_B$ at room T.\cite{JPhysCM14p6509y2002}

\begin{table}[b]
\caption{\label{table1}
$a$ and $c$, $c/a$, cell $V$, and M$\,(\mu_B$/MnBi) of LTP-MnBi from experiment and our (or former$^a$) DFT results.}
\begin{ruledtabular}
\begin{tabular}{lllrll}
 $a\,$(\AA) & $c\,$(\AA) & $c/a$ & $V\,$(\AA$^3$) & $\mu_B$ & Ref. \\
\hline
4.2827 & 6.1103 & 1.4267 & 97.0574  & 4.18 & \cite{JPhysCM14p6509y2002} @10~K\\
4.286  & 6.126	& 1.4293 & 97.4567	&  & \cite{JMMM115p66y1992} \\
4.28	& 6.11	& 1.427	& 96.9303	&  & \cite{PSSA34p553y1976} \\
4.305	& 6.118	& 1.4211 & 98.1943	&  & \cite{PSSA30p251y1975} \\
4.285	& 6.113	& 1.4266 & 97.2046	&  & \cite{JAP43p2358y1974} \\
4.32	& 5.84	& 1.352	& 94.3867	&  & \cite{JAP39p5471y1968} \\
\hline				
4.3080	& 5.7398 & 1.3324 &  92.2554	& 3.455 & GGA \\
4.3625	& 6.1231 & 1.4036 &  100.9217	& 3.96 & GGA+U \\
\hline
4.170 & 5.755 & 1.3801 & 86.6659 & 4.01 & LMTO  \cite{PRB59p15680y1999} \\
4.26 & 6.05 & 1.420 & 95.0835 & 3.7 & ASM  \cite{JPhysFMetPhys15p2135y1985} \\
4.30 & 6.12 & 1.423 & 97.9984 & 3.50 & LCAO \cite{JapJAP46p3455y2007} \\
\end{tabular}
\end{ruledtabular}
\footnotetext[1]{Note: $a$ and $c$ were fixed in \onlinecite{PRB59p15680y1999}, \onlinecite{JPhysFMetPhys15p2135y1985}, and \onlinecite{JapJAP46p3455y2007}.}
\end{table}

{\par}
While the GGA+U better describes strongly-correlated systems, like MnBi, there still remains a small systematic DFT error in the lattice constants, 
arising from the approximation in the exchange-correlation functional (which introduces a small shift in pressure, but not in the curvature of the total energy). 
Notably, the measured lattice constants differ by 1\%, e.g., at T=$50\,$K, $c$=$6.05\,${\AA} \cite{APL99p082505y2011} and  $6.11\,${\AA}. \cite{SciTechnolAdvMater9p024204y2008} 
The MAE (Fig.~\ref{figMAE}) is small and very sensitive to ($a,c$).
For proper comparison, we plot in Fig.~\ref{figMAE} both the measured ($a,c$) and those shifted by 0.8\% 
to account for a DFT bias in the GGA+U lattice constants for a given alloy. 

{\par} {\em Comparison to previous DFT calculations:}
Without the Hubbard U correction, GGA gives M(0) of $3.455\, \mu_B$ and distorts the cell,  underestimating its volume (Table 1).  
For comparison, previous DFT results are 3.50, \cite{JapJAP46p3455y2007} 3.49,  \cite{JPhysCM14p6509y2002} and $3.52\, \mu_B$.  \cite{JPhysFMetPhys15p2135y1985}
Fixing $a$ to $4.170$~{\AA } and $c$ to $5.755$~{\AA } gives a total moment of $4.01\, \mu_B$ in the full-potential LMTO, \cite{PRB59p15680y1999} while fixing $a$ to $4.26$~{\AA } and $c$ to $6.05~${\AA } gives  $3.7\, \mu_B$ 
in augmented spherical methods (ASM).\cite{JPhysFMetPhys15p2135y1985}
Magnetization of MnBi increases with volume. The calculated lattice constants, volume, and magnetization increase with the value of (U$-$J). 

{\par} {\em Results and Discussion:}
Around equilibrium, E($\Delta a$,$\Delta c$)  looks like a flat-bottom canoe, canted from a constant volume direction towards $c$ (Fig.~\ref{figEnergy}). 
Because the energy penalty for changing $c$ by 0.5\% is close to zero, 
even low-energy defects can alter $c$, and 
any value of $c$ within that range is accessible in experiment below $100\,$K.
Indeed, this predicted behavior of $c$ with chaotic amplitude within $\sim$$0.5$\% is observed. \cite{SciTechnolAdvMater9p024204y2008, APL99p082505y2011}

{\par } Below $6~me$V (70~K), E($\Delta a$,0) in Fig.~\ref{figEa} is symmetric with $\mbox{E}(+\Delta a,0)=\mbox{E}(-\Delta a,0)$, and can be well described by a parabola $\mbox{E}(a)=\frac{1}{2} \bar{m} \omega^2 (a-a_0)^2 $, where the unit cell mass is $\bar{m} = 2(m_{Mn}+m_{Bi})=527.836~$a.m.u., and  $\omega = 1.2 \cdot 10^{13}\,\mbox{s}^{-1} $ is harmonic  frequency for vibrations along $a$. 
Quantization of this potential results in a descrete  spectrum with the equidistant levels 
$  E_{n} = \hbar \omega ( n + \frac{1}{2} ) $, with 
$ \hbar \omega = 7.7\,m$eV ($90\,$K). 
Due to the symmetric potential and absence of vibrational excitations, there is no thermal expansion along $a$ at $\mbox{T}<70\,$K.

{\par} Above $9~me$V ($100\,$K), E($\Delta a$,0) is asymmetric with $\mbox{E}(+\Delta a,0)<\mbox{E}(-\Delta a,0)$. 
It can be approximated by a cubic polynomial 
$\mbox{E}(a) = E_0 + \epsilon_2 \Delta a^2 + \epsilon_3 \Delta a^3 $, with
$\epsilon_2 = 3.8\,$eV/{\AA}$^2$ and $\epsilon_3 = -2.1\,$eV/{\AA}$^3$. 
This fit has $\chi^2 = 6 \cdot 10^{-7}$, RMS relative error of $7.9 \cdot 10^{-6}$, 
and Theil U coefficent of $7.8 \cdot 10^{-6}$. 
For $N = 4$ ions per unit cell, our theoretical estimate of the linear thermal expansion coefficient ($\alpha_a = \frac{1}{a} \frac{da}{dT} \approx -\frac{1}{a} N k \frac{\epsilon_3}{\epsilon_2^2} $) is $1.153 \cdot 10^{-5}\,\mbox{K}^{-1}$, in agreement with experiment, \cite{SciTechnolAdvMater9p024204y2008} i.e., $1.168 \cdot 10^{-5}\,\mbox{K}^{-1}$.

{\par} Hence, the potential energy surface in Fig.~\ref{figEnergy} predicts no thermal expansion along $a$ at low $\mbox{T}<70\,$K,  and a positive expansion at higher T above 100~K, as observed. \cite{SciTechnolAdvMater9p024204y2008}

{\par} The spin reorientation in MnBi near $90\,$K was not fully understood in experiments. \cite{JETPLetters30p333y1979, JPhysCM14p6509y2002, JAP91p7866y2002}
Moreover, previous DFT calculations of MAE found the easy axis to be always in-plane (Table~3 in Ref.~\onlinecite{PRB59p15680y1999}).
We calculate dependence of the MAE on ($a,c$), and find that it is strongly affected by $a$ and very weakly by $c$, see Fig.~\ref{figMAE}. 
Thus, thermal expansion of $a$ causes the MAE to change from negative (in-plane oriented moments) to positive (moments oriented along the $c$-axis). 
This sign change causes a spin reorientation, experimentally observed around $90\,$K. \cite{JETPLetters30p333y1979,JPhysCM14p6509y2002,JAP91p7866y2002} 
Magnetic susceptibility has maximum at MAE=0, \cite{JETPLetters30p333y1979,APL79p1846y2001,JPhysCM14p6509y2002} 
when spins easily reorient along the external applied magnetic field.   
Coercivity is zero if $|\mbox{MAE}|<k\mbox{T}$, but increases with MAE at $\mbox{T}>180\,$K. \cite{JAP91p7866y2002}
Thus, dependence of MAE on ($a$,\,$c$) causes spin reorientation and explains the thermal behavior of magnetic susceptibility and coercivity. 


\begin{figure}
\includegraphics[width=60mm]{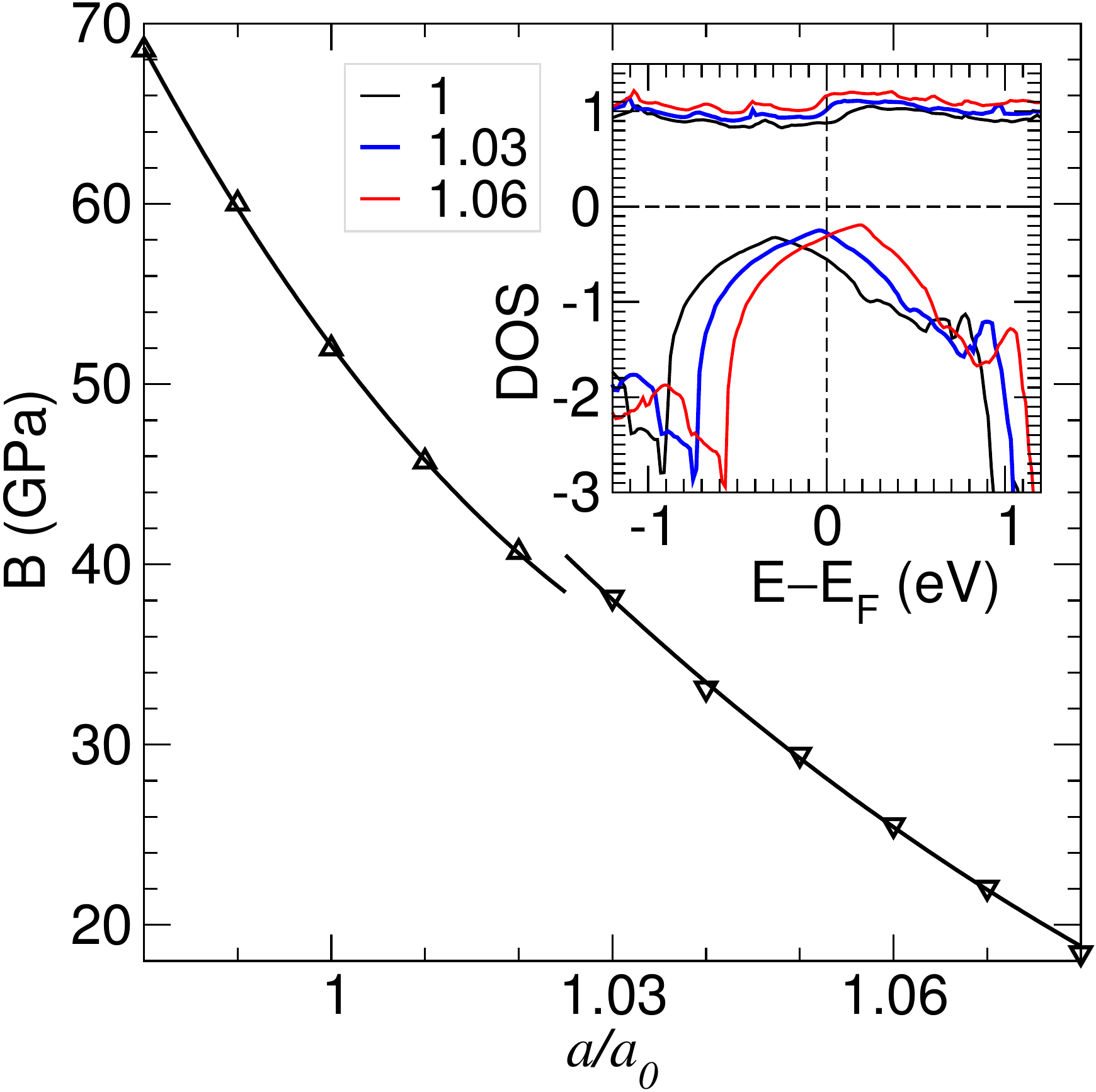}
\caption{\label{fig4Young} B vs. $a$ at isotropic expansion ($a/a_0=c/c_0$) relative to $a_0=4.2827${\AA
 } and $c_0=6.1103${\AA }. \cite{JPhysCM14p6509y2002}
(Inset) Spin DOS ({states}/[cell$\,\cdot\,$eV])  for 3 values of $a/a_0$. 
}
\end{figure}

{\par} Another consequence of the anomalous potential energy surface E$(a,c)$ is the observed kink in B near 39 GPa at $100\,$K, a long-standing puzzle.  \cite{JETPLetters30p333y1979}
We calculate B$=V d^2 \mbox{E}/dV^2$ from dependence of E$(a,c)$ on $V=ca^2 \sqrt{3}/2$ at isotropic expansion ($\Delta a=\Delta c$ in Fig.~\ref{figEnergy}).
We find that B versus $a$ (Fig.~\ref{fig4Young}) is not monotonic near B$=39\,$GPa, as observed. \cite{JETPLetters30p333y1979} 
This kink originates from a change in DOS at the Fermi level (Fig.~\ref{fig4Young}; see also Fig.~4 in Ref.~\onlinecite{JPhysFMetPhys15p2135y1985}).  
The Fermi level (E$_F$) is in a pseudo-gap, and the minimum in the minority-spin DOS passes through E$_F$ with thermal expansion of $a$; the DOS minimum corresponds to the $a$ at B=39 GPa (inset, Fig.~\ref{fig4Young}).

{\par} {\em Summary:}
We calculated dependence of the total energy and magneto-anisotropy energy on the lattice geometry for MnBi low-T phase. Our results explain the unusual structural and magnetic properties, heretofore unexplained.
From the potential energy surface, we reproduced and explained the observed anomalous behavior of (i) the lattice constants and (ii) bulk modulus. 
The calculated MAE changes sign with a small increase in $a$, which causes spin reorientation during thermal expansion.
(iii) The magnetic susceptibility has a maximum at MAE=0 (at spin reorientation).
(iv) Further increase of MAE with thermally expanding $a$ increases coercivity at $\mbox{T} > 180\,$K, where $|\mbox{MAE}|>k\mbox{T}$.

{\par}Due to its sensitivity on $a$, the MAE can be altered by temperature, pressure, doping, or interfacial strain. \cite{JAP109p07A740y2011,JAP107p09E303y2010,JPhysD46p095003y2013,JAP111p07E326y2012} 
To test whether doping can achieve a positive MAE at all temperatures, we performed preliminary, small-cell calculations that find that doping with selected (Ni, Rh, Pd, Ir) metals increases coercivity and stabilizes the spin orientation along $c$. 
More extensive calculations for $<$3\% cationic or anionic doped (substitutions and interstitials) cases are planned to establish the effects on lattice, magnetism, and stability.
Our understanding of the anomalous magneto-structural behavior offers an opportunity to develop improved MnBi-based permanent magnets.

This work is supported by the U.S. Dept. of Energy ARPA-E (REACT 0472-1526). Some methods were developed under support by the Office of Basic Energy Science, Division of Materials Science and Engineering. Ames Laboratory is operated for the U.S. DOE by Iowa State University under contract DE-AC02-07CH11358. We thank our REACT team and F.J. Pinski for useful discussions.

\end{document}